\documentclass[aps,prb,twocolumn,showpacs,superscriptaddress,floatfix]{revtex4-2}
\usepackage{dcolumn}
\usepackage{amsmath}
\usepackage{graphicx,epsfig,psfrag}
\usepackage{amssymb}
\usepackage{natbib}
\usepackage{url}
\usepackage[breaklinks=true]{hyperref}
\usepackage{mathtools}
\usepackage{subfigure}
\usepackage{color}
\hypersetup{
        colorlinks=true,
}
\usepackage{bookmark}
\usepackage{epic}

\usepackage{txfonts}
\definecolor{black}{rgb}{0.0, 0.0, 0.0}
\definecolor{red}{rgb}{1,0,0}
\definecolor{blue}{rgb}{0,0,1}
\definecolor{darkgreen}{rgb}{0,0.5,0}
\definecolor{forestgreen}{rgb}{0.13, 0.55, 0.13}

\newcommand{\hn}[1]{{\color{black}{#1}}}
\newcommand{\tac}[1]{{\color{black}{#1}}}

\begin{document}
\title{Self-energy method for time-dependent spectral functions of the Anderson impurity model within the time-dependent numerical renormalization group approach
}
\author{H. T. M. Nghiem}
\affiliation{
Phenikaa Institute for Advanced Study, Phenikaa University, 12116 Hanoi, Vietnam}
\author{T. A. Costi}
\affiliation
{Peter Gr\"{u}nberg Institut and Institute for Advanced Simulation, 
Research Centre J\"ulich, 52425 J\"ulich, Germany}
\begin{abstract}
  The self-energy method for quantum impurity models expresses the correlation part of the self-energy in terms of the ratio of two Green's functions and allows for a more accurate calculation of equilibrium spectral functions than is possible directly from the one-particle Green's function [Bulla {\it et al.}, J.  Phys.: Condens. Matter  {\bf 10}, 8365 (1998)], for example, within the numerical renormalization group method. In addition, the self-energy itself is a central quantity required in the dynamical mean field theory of strongly correlated lattice models. Here, we show how to generalize the self-energy method to the time-dependent
  situation for the prototype model of strong correlations, the Anderson impurity model. We use the equation of motion method to obtain closed expressions for the local Green's function in terms of a time-dependent correlation self-energy, with the latter being given as a ratio of a one-particle time-dependent Green's function and a \hn{higher-order correlation} function. We benchmark this self-energy approach to time-dependent spectral functions against the direct approach within the  time-dependent numerical renormalization group method. The self-energy approach improves the accuracy of time-dependent spectral function calculations, and the closed-form expressions for the Green's function allow for a clear picture of the time-evolution of
  spectral features at the different characteristic time-scales. The self-energy approach is of potential interest also for other quantum impurity solvers for real-time evolution, including time-dependent density matrix renormalization group and continuous-time quantum Monte Carlo techniques. 
\end{abstract}

\date{\today}

\maketitle

\section{Introduction}\label{sec:intro}
Understanding the time-evolution and nonequilibrium dynamics of quantum impurity systems is relevant to diverse areas of physics, including
transport through quantum dots \cite{HanHeary2007,Oguri2018}, time-resolved spectroscopy of correlated systems within nonequilibrium dynamical mean field theory \cite{Freericks2006,Aoki2014,Randi2017} and quantum impurities in optical lattices \cite{Knap2012,KanaszNagy2018}. A large number of methods have been
developed to address time-evolution in  these systems, including functional renormalization group \cite{Kennes2012a},  time-dependent numerical renormalization group (TDNRG) 
\cite{Anders2005,Anders2006,Nghiem2014a,Nghiem2014b}, time-dependent density matrix renormalization group (TD-DMRG) \cite{Daley2004,White2004}, flow equation \cite{Lobaskin2005}, and continuous-time quantum Monte Carlo (CT-QMC) \cite{Gull2011b}. In this paper we focus on the TDNRG approach and generalize the self-energy approach to equilibrium spectral functions within the numerical renormalization group (NRG) \cite{Bulla1998} to time-dependent situations within the TDNRG approach.

In previous work \cite{Nghiem2017,Nghiem2020}, we demonstrated the ability of the TDNRG method \cite{Anders2005,Anders2006} in calculating dynamical quantities for the time-dependent Anderson impurity model. 
In particular, the time-dependent spectral function $A(\omega,t)$ was investigated using several
different definitions of the reference time $t$ describing the transient dynamics to the long-time limit $t\to\pm \infty$.
More specifically,  spectral functions with reference times $T=t_1$, $T=t_2$ and $T=(t_1+t_2)/2$ (average, or Wigner time) were considered, with $\omega$ being the frequency resulting from a Fourier transform on the relative time difference $\tau=t_1-t_2$ of the corresponding retarded two-time Green's function $G(t_1,t_2)$. Different definitions are of relevance to either transport through quantum dots \cite{Jauho1994} or time-resolved photoemission spectroscopy \cite{Freericks2009,Randi2017}. The calculations made manifest how the charge and spin fluctuation time scales
of the Anderson model appear in the time-resolved spectral function and in the time-resolved photoemission intensity following a quench of the local level from an initial to a final state position and within the Kondo regime.
For example, the satellite peak of the local level starts to change its position as a result of the quench on a time scale corresponding to the coupling energy ($\Delta$) between the impurity and the conduction electrons ($\hbar/\Delta$), and the Kondo resonance, while starting to form already at short times, only fully develops on a time scale corresponding to $\hbar/k_{\rm B}T_{\rm K}$, where $k_{\rm B}T_{\rm K}$ is the Kondo scale \cite{Nordlander1999,Nghiem2017,Nghiem2020}.

The main advantages of the TDNRG for time-resolved spectral functions, are that it is non-perturbative, capturing correctly both low- and high-energy scales, applies to infinite times (in contrast to CT-QMC \cite{Gull2011b} and TD-DMRG\cite{Daley2004,White2004}), and yields time-dependent spectral functions on the real frequency axis and at zero or finite temperature. One disadvantage of the TDNRG is that the use of a Wilson chain results in imperfect thermalization in the long-time limit \cite{Rosch2012,Guettge2013,Nghiem2020}.

While the TDNRG provides numerical results that give an overall correct picture of the time-evolution of the spectral function of the Anderson impurity model, further improvements are desirable, particularly if one aims to apply the TDNRG approach as an impurity solver within nonequilibrium dynamical mean field theory \cite{Freericks2006,Aoki2014}.
For equilibrium (time-independent) systems, one such improvement, the self-energy method for calculating spectral functions \cite{Bulla1998}, has proven particularly useful. Within this approach, the spectral function $A(\omega)=-\mathrm{Im}[G(\omega)]/\pi$ of the local Green's function $G(\omega)$ is determined not via the NRG through the usual Lehmann representation \cite{Sakai1989,Costi1994,Hofstetter2000,Peters2006,Weichselbaum2007}, but indirectly via the NRG by first calculating the correlation self-energy $\Sigma(\omega)$ as a ratio of a one-particle Green's function and \hn{a higher-order correlation function} \cite{Bulla1998}, and then using this, together with the one-particle broadening function  $\Gamma(\omega)$ as input to a spectral function calculation,
\begin{align}
  A(\omega) & =-\frac{1}{\pi}\mathrm{Im}\left[\frac{1}{\omega -\varepsilon_d - \Gamma(\omega) - \Sigma(\omega)}\right],\label{eq:A-from-sigma}
\end{align}
where $\varepsilon_d$ is the local level position in the Anderson model. This approach has a number of advantages over the direct calculation. Thus, the true width of the satellite peaks in $A(\omega)$ is better described, and the Friedel sum-rule is satisfied essentially exactly. In addition, the correlation self-energy, calculated as a ratio of a one-particle Green's function and a \hn{higher-order correlation} function, is also improved over that obtained directly from the Green's function $G(\omega)$ by inversion \cite{Bulla1998}. For these reasons, in this paper we are interested in deriving an analogous representation for  the time-dependent case, in which the local Green's function is expressed in terms of a time-dependent correlation self-energy.

So far, studies on nonequilibrium systems have been largely based on the work of  Kadanoff and Baym \cite{Kadanoff1962}, in which the equation of motion (EOM) is used to derive expressions for the one-particle Green's function. By using the Dyson relation, the one-particle Green's functions are the solutions of integro-differential equations which include the self-energy functions. The latter equations are equivalent to the Keldysh formulation with the integral along the Keldysh contour \cite{Keldysh1965,vanLeeuwen2006}. These works are useful in many contexts, for example, in perturbative approaches to the non-equilibrium problem \cite{Stefanucci2013} and in extending impurity solvers to nonequilibrium for applications  within nonequilibrium dynamical mean field theory \cite{Aoki2014}.

In this paper, instead of following the Keldysh-Baym-Kadanoff formulation, we start with the EOM of the two-time  Green's functions, in which the one-particle Green's function is expressed  in terms of the \hn{higher-order correlation} function \cite{Zubarev1960}. This is motivated by the fact that non-perturbative methods, e.g., the NRG \cite{Bulla2008} and the continuous-time quantum Monte Carlo method \cite{Gull2011b},  can easily calculate also the \hn{higher-order correlation} function.
By applying the Fourier transformation on the relative time $\tau$, we have the Green's function expressed as the solution of an ordinary differential equation (ODE) where the time-dependent self-energy function is the ratio between the one-particle Green's function and the \hn{higher-order correlation} function.
From this, we obtain the analytic form of the time-dependent Green's function.
For the noninteracting case, we calculate directly the spectral function from the analytic form, while, in the interacting case, the spectral function is calculated with the time-dependent self-energy extracted from the TDNRG. The comparison between the spectral functions in the noninteracting and interacting cases allows for a more detailed understanding of the origins of spectral features.

The outline of the paper is as follows. Section~\ref{sec:formalism} presents the formalism, starting with a description of the model and the Green's function considered (Sec.~\ref{subsec:model+gf}), followed by the equations of motion for the latter (Sec.~\ref{subsec:eom+gf}), the transformation of these to center-of-mass and relative-time coordinates and the solutions of these equations, resulting in closed expressions for the Green's functions at positive and negative times (Sec.~\ref{subsec:com+solution}). 
For the noninteracting case, explicit analytic results for these Green's functions are obtained (Sec.~\ref{subsec:noninteracting-solution}). 
In Sec.~\ref{sec:results} we present numerical results using the above formalism, starting with the time-dependent self-energies calculated either directly within the TDNRG approach or within the above formalism.  We analyze the respective differences within the two approaches in Sec.~\ref{subsec:self-energy-errors}. Section~\ref{subsec:compare-noninteracting+interacting} considers a level quench on both the interacting and noninteracting model and compares their respective time-dependent spectral functions. An approximate calculation of elastic and inelastic scattering rates is also provided and used to discuss the degree of validity of the Friedel sum-rule at various times. We conclude with a summary and outlook in Sec.~\ref{sec:conclusion}. Technical appendices on the wide-band limit, used in obtaining the closed expressions for the Green's functions as solutions of the ODEs in Sec.~\ref{subsec:com+solution}, and the details of the ODE solvers used in the calculations are given in Appendixes~\ref{sec:wbl} and \ref{sec:ode}. \hn{A detailed comparison between the results of the analytic expressions and the TDNRG method for the noninteracting case is shown in Appendix~\ref{sec:noninteracting}}.

\section{Formalism}
\label{sec:formalism}
\subsection{Model}
\label{subsec:model+gf}
We consider the time-dependent Anderson impurity Hamiltonian
\begin{align}
  H(t)&=\sum_{\sigma}\varepsilon_d(t)n_{d\sigma}+U(t)n_{d\uparrow}n_{d\downarrow}+\sum_{k\sigma}\varepsilon_kc^+_{k\sigma}c_{k\sigma}\nonumber\\
  &+\sum_{k\sigma}V_{kd}(t)(c^+_{k\sigma}d_{\sigma}+{\rm H.c})\label{eq:AIM},
\end{align}
where $\varepsilon_d(t), U(t)$ and $V_{kd}(t)$ are the time-dependent local level, Coulomb repulsion and hybridization matrix element respectively.  The retarded two-time Green's function that we are interested in is defined as
\begin{align}
  G_{BC}(t_1,t_2)=-i\Theta(t_1-t_2)\langle\{\hat{B}(t_1),\hat{C}(t_2)\}\rangle,\label{eq:two-time-gf}
\end{align}
where $B$ and $C$ can be any local operators, although for our purposes in this paper we shall take $B=d_{\sigma}=C^{\dagger}$.

\subsection{Equation of motion}
\label{subsec:eom+gf}
The time-evolution of an operator is given by the equation
$i\frac{\partial A(t)}{\partial t}=[A(t),\hat{H}(t)]$.
Using the Hamiltonian defined in Eq.~(\ref{eq:AIM}) and the definition of the retarded Green's function in Eq.~(\ref{eq:two-time-gf}), we have the EOM of this
Green's function with respect to  the first time $t_1$,
\begin{align}
i\frac{\partial G_d^{\sigma}(t_1,t_2)}{\partial t_1}=&\delta(t_1-t_2)+\varepsilon_d(t_1)G_d^{\sigma}(t_1,t_2)+U(t_1)F^{\sigma}_{d}(t_1,t_2)\nonumber\\ &+\sum_k V_{dk}(t_1)G_{kd}^{\sigma}(t_1,t_2)\label{eq:Gre1}
\end{align}
where the \hn{higher-order correlation} function $F^{\sigma}_{d}(t_1,t_2)$ appearing on the right-hand side of (\ref{eq:Gre1}) is given by,
\begin{align}
F^{\sigma}_{d}(t_1,t_2)=&-i\Theta(t_1-t_2)\langle \{[d_{\sigma}d^+_{\bar{\sigma}}d_{\bar{\sigma}}](t_1),d_{\sigma}^+(t_2)\}\rangle\label{eq:F}.
\end{align}
Similarly, the EOM of the retarded Green's function with respect to  the second time $t_2$ reads
\begin{align}
i\frac{\partial G_d^{\sigma}(t_1,t_2)}{\partial t_2}=&-\delta(t_1-t_2)-\varepsilon_d(t_2)G_d^{\sigma}(t_1,t_2)-U(t_2)\tilde{F}^{\sigma}_{d}(t_1,t_2)\nonumber\\ &-\sum_k V_{kd}(t_2)G_{dk}^{\sigma}(t_1,t_2)\label{eq:Gre2}
\end{align}
where the \hn{higher-order correlation} function $\tilde{F}^{\sigma}_{d}(t_1,t_2)$ on the right-hand side of  Eq.~(\ref{eq:Gre2}) is given by
\begin{align}
\tilde{F}^{\sigma}_{d}(t_1,t_2)
  =-i\Theta(t_1-t_2)\langle \{d_{\sigma}(t_1),[d^+_{\sigma}d^+_{\bar{\sigma}}d_{\bar{\sigma}}](t_2)\}\rangle.\label{eq:F-tilde}
\end{align}
Finally, the equations of motion for the Green's function $G_{kd}^{\sigma}(t_1,t_2)=-i\Theta(t_1-t_2)\langle \{c_{k\sigma}(t_1),d_{\sigma}^+(t_2)\}\rangle$ and
$G_{dk}^{\sigma}(t_1,t_2)=-i\Theta(t_1-t_2)\langle \{d_{\sigma}(t_1),c_{k\sigma}^+(t_2)\}\rangle$
appearing on the right-hand side of (\ref{eq:Gre1}) and (\ref{eq:Gre2}) are,
  \begin{align}
i\frac{\partial G_{kd}^{\sigma}(t_1,t_2)}{\partial t_1}=&\varepsilon_{k\sigma}G_{kd}^{\sigma}(t_1,t_2)+V_{kd}(t_1)G_{d}^{\sigma}(t_1,t_2),\label{eq:Gre3}\\
i\frac{\partial G_{dk}^{\sigma}(t_1,t_2)}{\partial t_2}=&-\varepsilon_{k\sigma}G_{dk}^{\sigma}(t_1,t_2)-V_{dk}(t_2)G_{d}^{\sigma}(t_1,t_2).\label{eq:Gre4}
  \end{align}
  For the time-independent case $U(t)=U,\varepsilon_d(t)=\varepsilon_d$ and $V_{kd}(t)=V_{kd}$, the above Green's functions depend only on the time difference $\tau=t_1-t_2$,
  and the self-energy method of Bulla {\it et al.} \cite{Bulla1998} can be recovered by considering the Fourier transforms of the above equations,
  \begin{align}
    \omega G_{d}^{\sigma}(\omega) = & 1 + \varepsilon_d G_{d}^{\sigma}(\omega) + UF_{d}^{\sigma}(\omega) + \sum_{k}V_{kd}G_{kd}^{\sigma}(\omega)\\
    \omega G_{kd}^{\sigma}(\omega) = & \varepsilon_{k}G_{kd}^{\sigma}(\omega) + V_{kd}G_{d}^{\sigma}(\omega),
  \end{align}
which yield,
   \begin{align}
    G_{d}^{\sigma}(\omega) = & \frac{1}{\omega-\varepsilon_d -\Gamma(\omega) -\Sigma^{\sigma}(\omega)},\label{eq:gf-sigma}
  \end{align}
  with $\Sigma^{\sigma}(\omega)=UF_{d}^{\sigma}(\omega)/G_{d}^{\sigma}(\omega)$ being the equilibrium correlation self-energy and $\Gamma(\omega)=\sum_{k}|V_{kd}|^2/(\omega-\varepsilon_k)$ being the hybridization function \cite{Bulla1998}. As shown in Ref.~\onlinecite{Bulla1998}, by \hn{first evaluating $\Sigma^{\sigma}(\omega)$ from an NRG calculation of $F_{d}^{\sigma}$ and $G_{d}^{\sigma}$ (through their usual Lehmann representations) and then substituting the calculated $\Sigma^{\sigma}(\omega)$ and the known $\Gamma(\omega)$ back into Eq.~(\ref{eq:gf-sigma}),  a more 
accurate spectral function (\ref{eq:A-from-sigma})} is obtained than from an NRG calculation of $G_{d}^{\sigma}$ alone. The following section generalizes this approach to the time-dependent case.

\subsection{Center-of-mass and relative-time coordinates and sudden quench}
\label{subsec:com+solution}
In the previous study \cite{Nghiem2020}, we show that the Green's function defined with average time is relevant to the time-resolved photoemission spectroscopy observation; therefore, in this paper, we derive the analytic form of the Green's function with average time. The time transformation is defined such that $(t_1+t_2)/2=T$ and $t_1-t_2=\tau$.  Due to this transformation, we have the relations 
\begin{align}
\frac{\partial }{\partial t_1}+\frac{\partial }{\partial t_2}=\frac{\partial }{\partial T}&\label{eq:aveT1}\\
\frac{1}{2}\Big(\frac{\partial }{\partial t_1}-\frac{\partial }{\partial t_2}\Big)=\frac{\partial }{\partial \tau}\label{eq:aveT2}.
\end{align}
\begin{widetext}
In the following, we consider the system in response to  sudden quench given by $\varepsilon_d(t)=\theta(-t)\varepsilon^i_d+\theta(t)\varepsilon^f_d$, $U(t)=\theta(-t)U^i+\theta(t)U^f$ and $V_{kd}(t)=\theta(-t)V_{kd}^i+\theta(t)V_{kd}^f$. Using Eqs.~(\ref{eq:aveT1}) and (\ref{eq:aveT2}), we have   Eqs.~(\ref{eq:Gre1}-\ref{eq:Gre4})  equivalent to
\begin{align}
&\Big[i\frac{\partial}{\partial \tau}+i\frac{\partial}{2\partial T}-\varepsilon_d(T+\tau/2)\Big]G^{\sigma}_d(T,\tau)=\delta(\tau)+\sum_kV_{dk}(T+\tau/2)G^\sigma_{kd}(T,\tau)+U(T+\tau/2)F^\sigma_{d}(T,\tau),\label{eq:Gre1b}\\
&\Big[i\frac{\partial}{\partial \tau}-i\frac{\partial}{2\partial T}-\varepsilon_d(T-\tau/2)\Big]G^{\sigma}_d(T,\tau)=\delta(\tau)+\sum_kV_{kd}(T-\tau/2)G^\sigma
_{dk}(T,\tau)+U(T-\tau/2)\tilde{F}^\sigma_{d}(T,\tau),\label{eq:Gre2b}\\
&\Big[i\frac{\partial}{\partial \tau}+i\frac{\partial}{2\partial T}-\varepsilon_{k\sigma}\Big]G^{\sigma}_{kd}(T,\tau)=V_{kd}(T+\tau/2)G^\sigma_{d}(T,\tau),\label{eq:Gre3b}\\
&\Big[i\frac{\partial}{\partial \tau}-i\frac{\partial}{2\partial T}-\varepsilon_{k\sigma}\Big]G^{\sigma}_{dk}(T,\tau)=V_{dk}(T-\tau/2)G^\sigma_{d}(T,\tau)\label{eq:Gre4b}.
\end{align}
\tac{Solving Eqs.~(\ref{eq:Gre1b}) and (\ref{eq:Gre3b}) for positive times $T>0$, and Eqs.~(\ref{eq:Gre2b}) and (\ref{eq:Gre4b}) for negative times $T<0$, in the wide-band limit (see Appendix~\ref{sec:wbl}), we have}
\begin{align}
&\Big[\omega-{\varepsilon_d^f}-\Gamma^f(\omega)\Big]G^{\sigma}_d(T,\omega)+i\frac{\partial}{2\partial T}G^{\sigma}_d(T,\omega)
=1+{U^f}F^\sigma_{d}(T,\omega)\label{eq:ODE1}
\end{align}
at $T>0$, \tac{with $\Gamma^f(\omega)=-i\pi |V^f|^2\rho=-i\Delta^f$, and,}
\begin{align}
&\Big[\omega-{\varepsilon_d^i}-\Gamma^i(\omega)\Big]G^{\sigma}_d(T,\omega)-i\frac{\partial}{2\partial T}G^{\sigma}_d(T,\omega)
=1+{U^i}\tilde{F}^\sigma_{d}(T,\omega)\label{eq:ODE2}
\end{align}  
at $T<0$ \tac{with $\Gamma^i(\omega)=-i\pi |V^i|^2\rho=-i\Delta^i$.}
\tac{Equations~(\ref{eq:ODE1}) and (\ref{eq:ODE2}) are first-order inhomogeneous ordinary-differential equations with a boundary condition at $T=0$ given by adding Eqs.~(\ref{eq:ODE1}) and (\ref{eq:ODE2}),}
\begin{align}
\Big[\omega-\frac{\varepsilon_d^f+\varepsilon_d^i}{2}-\frac{\Gamma^f(\omega)+\Gamma^i(\omega)}{2}\Big]G^{\sigma}_d(T=0,\omega)
=1+\frac{U^f}{2}F^\sigma_{d}(T=0,\omega)+\frac{U^i}{2}\tilde{F}^\sigma_{d}(T=0,\omega).
\end{align}

\tac{We define the self-energy function for positive times $T>0$ by
$\Sigma^{\sigma}(T,\omega)=U^f F^{\sigma}_d(T,\omega)/G^{\sigma}_d(T,\omega)$ and for negative times $T<0$ by} $\tilde{\Sigma}^{\sigma}(T,\omega)=U^i \tilde{F}^{\sigma}_d(T,\omega)/G^{\sigma}_d(T,\omega)$. Then the solutions of Eqs.~(\ref{eq:ODE1}) and (\ref{eq:ODE2}) are as follows:
\begin{align}
G^{\sigma}_d(T>0,\omega)=e^{2i\int^T_{0}[\omega-\varepsilon_d^f-\Gamma^f(\omega)-\Sigma^{\sigma}(T_1,\omega)]dT_1}g(\omega)-2i\int_0^T e^{2i\int^{T}_{T_1}[\omega-\varepsilon_d^f-\Gamma^f(\omega)-\Sigma^{\sigma}(T_2,\omega)]dT_2}dT_1,\label{eq:solution1}
\end{align}
\begin{align}
G^{\sigma}_d(T<0,\omega)=e^{-2i\int^T_{0}[\omega-\varepsilon_d^i-\Gamma^i(\omega)-\tilde{\Sigma}^{\sigma}(T_1,\omega)]dT_1}g(\omega)+2i\int_0^T e^{-2i\int^{T}_{T_1}[\omega-\varepsilon_d^i-\Gamma^i(\omega)-\tilde{\Sigma}^{\sigma}(T_2,\omega)]dT_2}dT_1,\label{eq:solution2}
\end{align}
with
\begin{align}
g(\omega)=G^{\sigma}_d(T=0,\omega)=\Big[\omega-\frac{\varepsilon_d^f+\varepsilon_d^i}{2}-\frac{\Gamma^f(\omega)+\Gamma^i(\omega)}{2}-\frac{\Sigma^{\sigma}(T=0,\omega)+\tilde{\Sigma}^{\sigma}(T=0,\omega)}{2}\Big]^{-1}\label{eq:boundary}.
\end{align}

Equations (\ref{eq:solution1}) and (\ref{eq:solution2}) are the analytic forms of the Green's function at positive and negative times. So with the given 
self-energy functions, one may determine the Green's function via these equations 
\tac{using the numerical solvers presented in detail} in Appendix~\ref{sec:ode}. In the long-time limits, we have
\begin{align}
G^{\sigma}_d(T\to+\infty,\omega)=\Big[\omega-{\varepsilon_d^f}-{\Gamma^f(\omega)}-{\Sigma^{\sigma}(T\to+\infty,\omega)}\Big]^{-1},\label{eq:infinity}\\
G^{\sigma}_d(T\to-\infty,\omega)=\Big[\omega-{\varepsilon_d^i}-{\Gamma^i(\omega)}-{\tilde{\Sigma}^{\sigma}(T\to-\infty,\omega)}\Big]^{-1},\label{eq:initial}
\end{align}
in which \tac{the first equation gives the equilibrium final-state Green's function, while the second equation returns the initial-state Green's function.}
\subsection{Noninteracting case}
\label{subsec:noninteracting-solution}
In the noninteracting case, $U^i=U^f=0$ and then $\Sigma=\tilde{\Sigma}=0$, we 
\tac{obtain closed analytic expressions for the Green's functions from Eqs.~(\ref{eq:solution1}) and (\ref{eq:solution2})}
\begin{align}
G^{\sigma}_d(T>0,\omega)=\frac{1 -e^{2i[\omega-\varepsilon_d^f-\Gamma^f(\omega)]T}}{\omega-\varepsilon_d^f-\Gamma^f(\omega)}+  \frac{e^{2i[\omega-\varepsilon_d^f-\Gamma^f(\omega)]T}}{\omega-\frac{\varepsilon_d^f+\varepsilon_d^i}{2}-\frac{\Gamma^f(\omega)+\Gamma^i(\omega)}{2}}\label{eq:nonint1},
\end{align}
\tac{and,}
\begin{align}
G^{\sigma}_d(T<0,\omega)=\frac{1 -e^{-2i[\omega-\varepsilon_d^i-\Gamma^i(\omega)]T}}{\omega-\varepsilon_d^i-\Gamma^i(\omega)}+  \frac{e^{-2i[\omega-\varepsilon_d^i-\Gamma^i(\omega)]T}}{\omega-\frac{\varepsilon_d^f+\varepsilon_d^i}{2}-\frac{\Gamma^f(\omega)+\Gamma^i(\omega)}{2}}\label{eq:nonint2}.
\end{align}
Apparently, the time-dependent Green's function at positive times is a superposition between the Green's functions of the final-state and intermediate-state, 
while that at negative times is a superposition between the Green's functions of the initial-state and intermediate-state. 
The time-evolution factor involves the energy scales of either the \tac{initial- or the final-state} depending on whether the time lies before or after the quench. 
For time reference defined via $T=t_1$,  a similar form for the Green's function can be found elsewhere \cite{Jauho1994}, while the derivation for the definition with $T=t_2$ is easily carried out by following the steps in this section.
\end{widetext}

\section{Numerical results}
\label{sec:results}
Below we first compare numerical results for the self-energy obtained either directly from the TDNRG retarded Green's function (Sec.\ref{subsec:self-energy-errors}) or indirectly as a time-dependent correlation self-energy involving a ratio of a one-particle Green's function and a \hn{higher-order correlation} function and discuss the errors involved for the cases $T=0$ and $T=\pm \infty$. Results using the time-dependent correlation self-energy are then shown for all times. In Sec.~\ref{subsec:compare-noninteracting+interacting}, we use the time-dependent self-energies to calculate the time-dependent spectral function of the Anderson model
in response to a quench, comparing also with results for the noninteracting case. 
In order to benchmark our results against those from a direct calculation of the time-dependent spectral function \cite{Nghiem2020}, we shall consider the same symmetric quench as that used in Ref.~\onlinecite{Nghiem2020}, i.e.,
$\varepsilon_d(t) =\theta(-t)\varepsilon^i_d+\theta(t)\varepsilon^f_d$, with $\varepsilon^i_d =-0.015$ and $\varepsilon^f_d=-0.006$, $U(t)=\theta(-t)U^i+\theta(t)U^f$ with $U_i=0.03$ and $U_f=0.012$,
and constant \tac{equal hybridization functions in initial and final states $i\Gamma^{i}(\omega)=\Delta^i =i\Gamma^f(\omega)=\Delta^f=\Delta=\pi\rho V^2=0.001$}, where the half-bandwidth $D=1$ is the unit of energy. Therefore the Kondo temperature of the system is $T^{\rm i}_K=3\times 10^{-8}$ in the initial state and $T_{\rm K}=2.5\times 10^{-5}$ in the final state. In the following we shall show spectra as a function of $\omega/T_{\rm K}$,
for both interacting and noninteracting cases (for the noninteracting case only high energy peaks are present).

\subsection{Self-energy function}
\label{subsec:self-energy-errors}
\begin{figure}%
\includegraphics[width=0.48\textwidth]{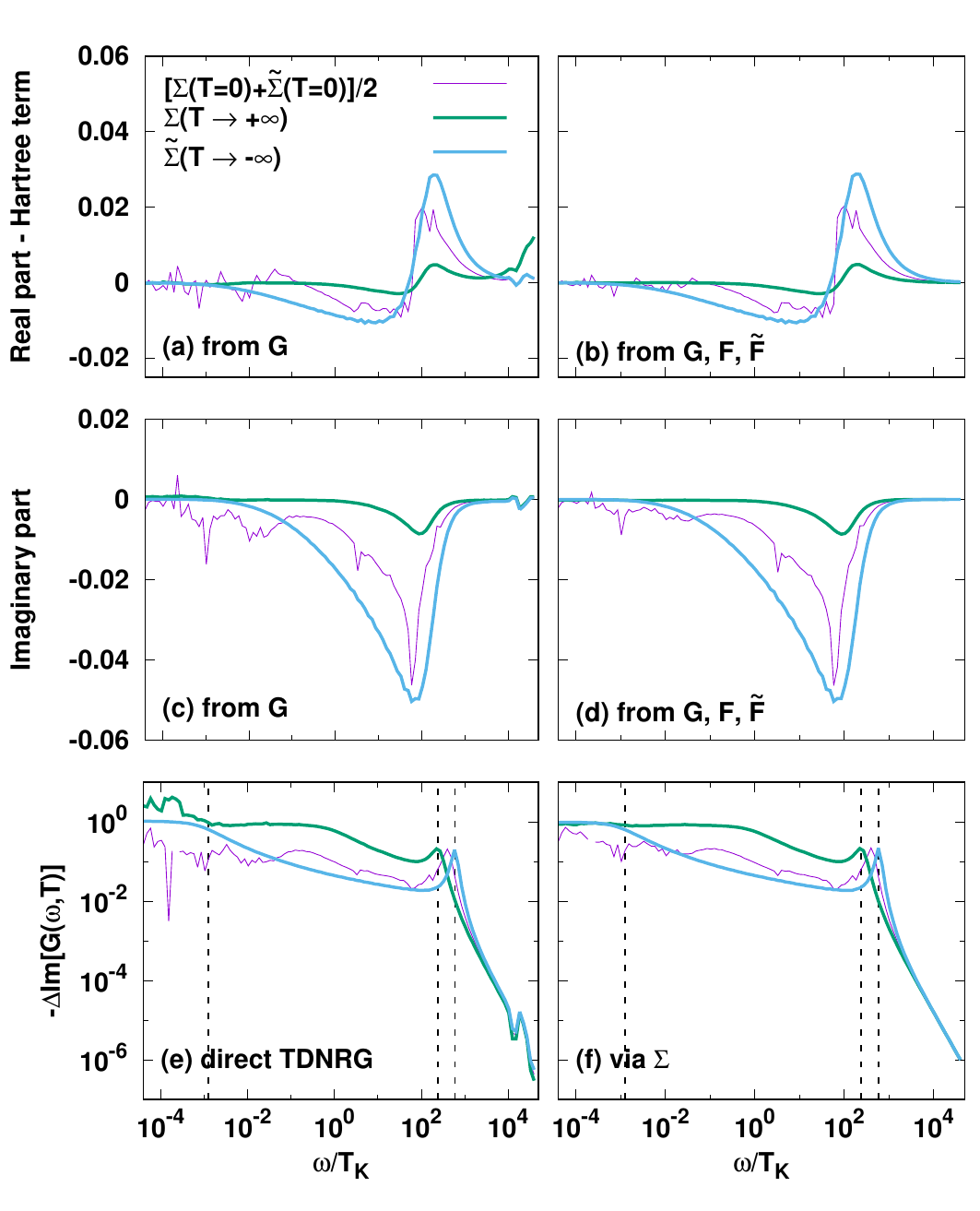}
\caption{The real and imaginary parts of the self-energy functions calculated only from $G$ in (a) and (c) and from $G$, $F$ and $\tilde{F}$ in (b) and (d)  at $T=0$, $+\infty$, and $-\infty$. {Note that the Hartree term $U(T)\langle n_{d\bar{\sigma}}(T)\rangle$ has been subtracted from the real parts of the self-energy}. In (e) and (f), the normalized imaginary-part of the Green's functions at $T=0$, $+\infty$, and $-\infty$ are calculated directly by TDNRG and through the self-energy functions shown in (b) and (d). The vertical dashed lines, from left to right, indicate the position of \hn{$T_{\rm K}^i$}, $|\varepsilon^{f}|=\frac{U^f}{2}$ and $|\varepsilon^{i}|=\frac{U^i}{2}$. The symmetric quench described at the beginning of Sec.~\ref{sec:results} is used.
The TDNRG calculations use as discretization parameter $\Lambda=4$, $z$ averaging \cite{Campo2005} with $N_z=32$ and a cutoff energy $E_{cut}=24$. \hn{The Lorentzian broadening scheme is used with the broadening width $\eta=b|E^m_{rs}|$ where $b=1/N_z$ and $E^m_{rs}$ is the excitation energy (the supplementary material of reference \cite{Nghiem2017}). The average time $T$ should not be confused as the temperature. All  calculations here, and in later figures, are at zero temperature.}
}\label{fig:compare3}%
\end{figure}

In Sec.~\ref{subsec:eom+gf}, closed expressions were derived for the time-dependent retarded Green's function, at positive and negative times, in terms of corresponding time-dependent self-energies.
  However, the self-energy functions themselves have no closed analytic form, but can be determined numerically.
  In this paper, we use the TDNRG for the numerical calculations.

We here first assess the accuracy of two different ways of extracting the self-energy functions, either (i) 
  by inverting Eqs.~(\ref{eq:solution1}) and (\ref{eq:solution2}) for a given $G^{\sigma}_d(T,\omega)$ to obtain the self-energies or  (ii)
  by directly evaluating them within TDNRG via the definitions $\Sigma^{\sigma}(T,\omega)=U^f F^{\sigma}_d(T,\omega)/G^{\sigma}_d(T,\omega)$ and $\tilde{\Sigma}^{\sigma}(T,\omega)=U^i \tilde{F}^{\sigma}_d(T,\omega)/G^{\sigma}_d(T,\omega)$. We also discuss how the two approaches affect the accuracy of the resulting one-particle Green's function. For simplicity, we only compare here the self-energy functions for times $T=-\infty$, $0$, and $+\infty$, since  at these times, the inversion in the approach (i) simply reduces to inverting the algebraic equations~(\ref{eq:boundary}-\ref{eq:initial}).

Considering a value calculated by TDNRG having a form of $A=\bar{A}+\delta A$ with $\bar{A}$ being the expected value and $\delta A$ being the absolute error, we have the error of the self-energy function calculated in the first approach as
\begin{align}
\delta \Sigma_1= \frac{1}{\bar{G}}-\frac{1}{{G}}= \frac{\delta G}{G\bar{G}}\label{eq:error1}.
\end{align}
while in the second way the error is given by
\hn{\begin{align}
\delta \Sigma_2= \Sigma - \bar{\Sigma}= U\frac{F}{G}-U\frac{\bar{F}}{\bar{G}}
=U\frac{\delta F \bar{G}-\delta G \bar{F}}{G\bar{G}}\label{eq:error2}, 
\end{align}
The ratio of the errors in the self-energy calculated in the two approaches $\delta \Sigma_2/\delta\Sigma_1$ can be expressed in terms of relative errors of $F$ and $G$ as
\begin{align}
\frac{\delta \Sigma_2}{\delta\Sigma_1}= U\bar{F}\frac{\frac{\delta F}{\bar{F}}-\frac{\delta G}{\bar{G}}}{\frac{\delta G}{\bar{G}}}\label{eq:error-ratio}.
\end{align}
Since $F$ and $G$ are calculated in the same way in TDNRG, their relative errors are similar, hence the numerator in Eq.~(\ref{eq:error-ratio}), which subtracts comparable relative errors, can be considerably smaller than the relative error in the denominator [note that  the prefactor, $U\bar{F}$, in (\ref{eq:error-ratio}) is of $O(1)$ for $U\to\infty$, as can be verified, for example, in the zero bandwidth limit $\Delta \to 0$].}
Therefore, in the subtraction appearing in the numerator in Eq.~(\ref{eq:error2}), the error in the second approach  can be significantly reduced \cite{Bulla1998}. Hence, using the self-energy function calculated by the second way to calculate $G^{\sigma}_d(T,\omega)$ may also give more accurate results
than calculating $G^{\sigma}_d(T,\omega)$ directly within the TDNRG.

For a more detailed understanding, we show the self-energy functions calculated in the two approaches at the different times in Fig.~\ref{fig:compare3}. In Figs.~\ref{fig:compare3}(a) and \ref{fig:compare3}(c), the self-energies calculated by the first approach via $G$ are shown, whereas in Figs.~\ref{fig:compare3} (b) and \ref{fig:compare3}(d) those calculated by the second approach via  $G$, $F$, and $\tilde{F}$ are shown\footnote{\label{Note1}Note that the Hartree terms, given by $(U^i+U^f)/4$, $U^f/2$, and $U^i/2$ respectively, have been subtracted out from the real parts of the self-energy  at $T=0$, $+\infty$, and $-\infty$. The values of these Hartree terms follow from \unexpanded{$\langle n_{d\bar{\sigma}}(T) \rangle= 0.5$} since the system is particle-hole symmetric both before and after the quench.}. We see that the results of the first approach show deviations from the expected value of zero at large frequencies, which are similar for the different times considered. In contrast, the results from the second approach show the convergence to the correct value at the large frequencies. The results are consistent with the error analysis from Eqs.~(\ref{eq:error1}) and (\ref{eq:error2}). At low frequencies, the difference between the self-energy functions at $T=-\infty$ calculated via the two approaches is not visible, and the same holds for the self-energy functions at $T=+\infty$. In contrast,
differences between the self-energy functions calculated in the two approaches become evident at finite times, illustrated here for $T=0$.

In Fig.~\ref{fig:compare3}(e), we show the imaginary parts of the Green's functions calculated directly by the TDNRG [whose inversion, as discussed in the first approach above, yields the self-energy function shown in Fig.~\ref{fig:compare3}(a) and \ref{fig:compare3}(c)]. At high frequencies $|\omega|\gg U^{i,f}/2$, spurious peaks in the Green's function are seen,
which are directly related to the above mentioned deviations of the self-energy functions at these frequencies.
In Fig.~\ref{fig:compare3}(f), we show the imaginary parts of the Green's functions calculated from the self-energy functions shown in Fig.~\ref{fig:compare3}(b) and (d) obtained within the second approach.
In Fig.~\ref{fig:compare3}(f), the peaks observed in Fig.~\ref{fig:compare3}(e) do not appear, because the error of the self-energy functions at the high frequencies is reduced according to Eq.~(\ref{eq:error2}).
At low frequencies, the difference between the Green's functions calculated by the TDNRG directly and calculated via the self-energy functions is evident. At $T=-\infty$ and $+\infty$, the Friedel sum rule is better satisfied in the latter calculations than in the former [compare Figs.~\ref{fig:compare3}(e) and \ref{fig:compare3}(f)]. \hn{In particular, the violation of the sum rule of the spectral function in Fig.~\ref{fig:compare3}(f) is $2\%$ at $T=-\infty$ and $10\%$ at $T=+\infty$}. A ``violation'' of this sum rule is evident at short times $T=0$, a point we return to in the next section.
  
\begin{figure}%
\includegraphics[width=0.48\textwidth]{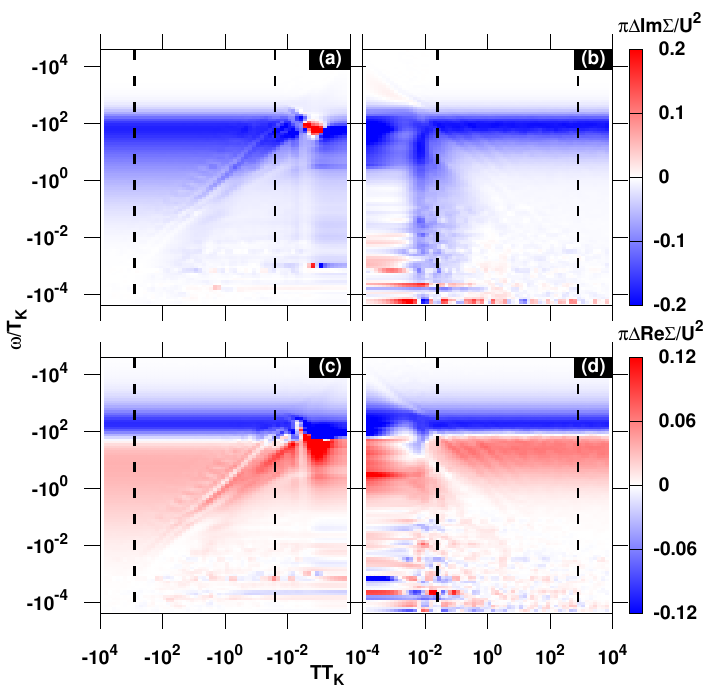}
\caption{(a) and (b) The imaginary part of the self-energy: $\mathrm{Im}[\tilde{\Sigma}^{\sigma}]$  at $T<0$ and $\mathrm{Im}[{\Sigma}^{\sigma}]$ at $T>0$, normalized by multiplying with $\pi\Delta/(U^i)^2$ and $\pi\Delta/(U^f)^2$ respectively. { (c) and (d) show the 
  real part of the self-energy with the Hartree term subtracted out, i.e.,} $\mathrm{Re}[\tilde{\Sigma}^{\sigma}]-\langle n_{d{\sigma}}(T)\rangle U^i$  at $T<0$ and $\mathrm{Re}[{\Sigma}^{\sigma}]-\langle n_{d\sigma}(T)\rangle U^f$ at $T>0$. Since the system is kept at particle-hole symmetry we have that $\langle n_{d{\sigma}}(T)\rangle=0.5$. \hn{Dashed vertical lines indicate $T=\pm 1/\Delta\approx\pm 1/40T_{\rm K}$ and $T=\pm 1/T_{\rm K}^i\approx\pm 800/T_{\rm K}$}. Quench and TDNRG parameters are as in Fig.~\ref{fig:compare3}
}
\label{fig:compare2}%
\end{figure}

We can see that the self-energy function calculated as a ratio between the one-particle Green's function and the \hn{higher-order correlation} function shows higher accuracy than that from the calculation via the one-particle Green's function only. Consequently, the one-particle Green's function calculated from the self-energy is more precise than that calculated directly via the TDNRG.

In Fig.~\ref{fig:compare2}, we show the self-energy functions calculated as the ratio of $F(\omega,T)$ and $G(\omega,T)$ for all times, both positive and negative. \hn{The self-energy at positive frequencies is not shown in the figure, however, since the system is kept at particle-hole symmetry, the self-energy function satisfies $\mathrm{Im}[\Sigma^{\sigma}(\omega)]=\mathrm{Im}[\Sigma^{\sigma}(-\omega)]$ and $\mathrm{Re}[\Sigma^{\sigma}(\omega)]=-\mathrm{Re}[\Sigma^{\sigma}(-\omega)]$.} We see that, at $\omega\to \pm\infty$, the self-energy function converges to the correct value, which is $\mathrm{Im}[\Sigma^{\sigma}]=0$ and $\mathrm{Re}[\Sigma^{\sigma}]-U(T)\langle n_{d\bar{\sigma}}\rangle=0$. One also observes the additional sign change of the real part of the self-energy at finite frequency (in addition to the usual one at $\omega=0$), a characteristic feature of strong correlations \cite{Bulla1998}. 
The imaginary part of the self-energy is generally negative for long times and for not too low frequencies [Fig.~\ref{fig:compare2}(a)]. Positive regions are observed for time-scales and frequencies where spectral weight is being rearranged (see next section), such as for $|T|\lesssim 1/\Delta$ when the local level is shifting between the initial and final state positions, or on longer time-scales and low energies $|\omega|\lesssim T_{\rm K}$ due to the buildup of the final-state Kondo resonance (or the destruction of the initial-state Kondo resonance at long negative times).

\subsection{Noninteracting case vs. interacting cases}
\label{subsec:compare-noninteracting+interacting}

\begin{figure}%
\includegraphics[width=0.48\textwidth]{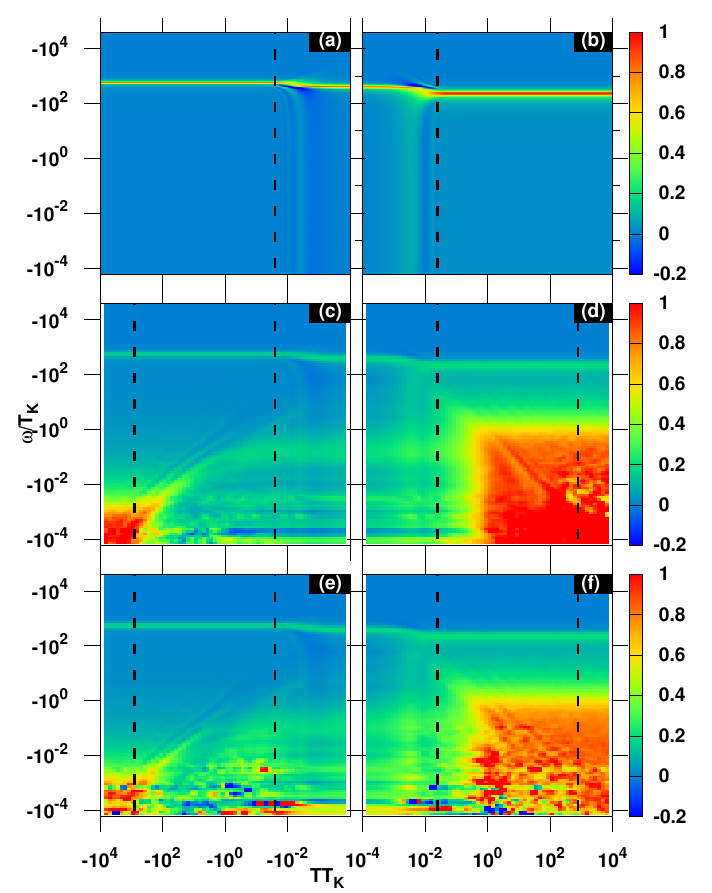}
\caption{(a) and (b) The normalized spectral function, $\pi\Delta A(\omega,T)$, in the noninteracting case calculated from the analytic equations (\ref{eq:nonint1}) and (\ref{eq:nonint2}).  (c) and (d) The normalized spectral function in the interacting case,  calculated directly from the TDNRG approach \hn{(from Ref.~\onlinecite{Nghiem2020})}. (e) and (f) The normalized spectral function in the interacting case calculated indirectly via the TDNRG using the self-energy function. \hn{Dashed vertical lines indicate $T=\pm 1/\Delta\approx\pm 1/40T_{\rm K}$ and $T=\pm 1/T_{\rm K}^i\approx\pm 800/T_{\rm K}$}. Quench and TDNRG parameters in (c)-(f) are as in Fig.~\ref{fig:compare3}}
\label{fig:compare1}%
\end{figure}

Using Eqs.~(\ref{eq:nonint1}) and (\ref{eq:nonint2}), we directly calculate the time-evolution of the spectral function, $A(\omega,T)=-\mathrm{Im}[G(\omega,T)]/\pi$, following a quench on the local level given by  $\varepsilon_d(T)=\theta(-T)\varepsilon^i_d+\theta(T)\varepsilon^f_d$ in the noninteracting case. In the wide-band limit, we have $\Gamma(\omega)=\Gamma^i(\omega)=\Gamma^f(\omega)=-i\Delta$ in which $\Delta$ is real and positive, see Appendix~\ref{sec:wbl}. 
In Figs.~\ref{fig:compare1}(a) and \ref{fig:compare1}(b), we show the spectral function when $\varepsilon^i_d=-0.015$, $\varepsilon^f_d=-0.006$, $\Delta^i=\Delta^f=\Delta=0.001$, and $U(T)=U^i=U^f=0$.  At infinite-negative time, the spectral function is that of the initial state with a well-defined peak at $\omega={\varepsilon^i_d}$, while, at $T=0^-$, the spectral function
shows the presence of an intermediate state with a  well-defined peak at the average energy $\omega=\frac{\varepsilon^i_d+\varepsilon^f_d}{2}$.
At the intermediate time $T$ close to $-1/\Delta$ (vertical dashed line), the contributions of both the initial and intermediate states to the spectral function are combined with the oscillatory function $e^{2i(\omega-\varepsilon^i_d)T}$ and the decay function $e^{-2\Delta |T|}$ [Eq.~ (\ref{eq:nonint2})]. Therefore we observe a spectral function strongly oscillating between positive and negative values.  At positive times, we recover the spectral function of the final state at $T\to\infty$ with a well-defined peak at the final state energy $\omega={\varepsilon^f_d}$. For times  $T$ close to $+1/\Delta$ (vertical dashed line) we observe the mixing of the final ($\omega={\varepsilon^f_d}$)  and the intermediate ($T\to 0^+$) state ($\omega=\frac{\varepsilon^i_d+\varepsilon^f_d}{2}$) in the spectral function, with the oscillation and decay functions as in Eq.~(\ref{eq:nonint1}). This results, again, in regions of positive and negative spectral density. Thus, as noted in previous work for the time-dependent spectral function of interacting systems [see Ref.~\onlinecite{Nghiem2020} and Fig.~\ref{fig:compare1}(c) and \ref{fig:compare1}(d)], negative regions of spectral density arise on time scales where the spectral weight is being strongly rearranged (for the noninteracting case, these correspond to times close to $T=\pm 1/\Delta$ when the peak in the spectral function is transitioning from $\varepsilon^i_d$ to $\frac{\varepsilon^i_d+\varepsilon^f_d}{2}$ and from $\frac{\varepsilon^i_d+\varepsilon^f_d}{2}$ to $\varepsilon^f_d$). \hn{We also compare the analytic results and the TDNRG results for the noninteracting case in Appendix~\ref{sec:noninteracting}. The overall TDNRG results are similar to the analytic results, however, there is a small difference in the magnitude and width of the high-energy spectral features associated with the local level position.}

To compare the spectral functions of the noninteracting and interacting cases, we show the spectral function in the interacting case with the same parameters as in the noninteracting case except that the quench for the interacting system also includes a Coulomb energy $U(T)=\theta(-T)U^i+\theta(T)U^f$ with $U^i=0.03$ and $U^f=0.012$. Figures~\ref{fig:compare1}(c) and \ref{fig:compare1}(d) show the time-dependent spectral function calculated by TDNRG from our previous work \cite{Nghiem2020}, while  Figs.~\ref{fig:compare1}(e) and \ref{fig:compare1}(f) show the spectral functions calculated in the approach of this paper via Eqs.~(\ref{eq:solution1}) and (\ref{eq:solution2}) with the self-energy functions given by $\Sigma(T,\omega)=U^f F^{\sigma}_d(T,\omega)/G^{\sigma}_d(T,\omega)$ and $\tilde{\Sigma}(T,\omega)=U^i \tilde{F}^{\sigma}_d(T,\omega)/G^{\sigma}_d(T,\omega)$, Fig. \ref{fig:compare2}. 

In both the time-dependent spectral functions calculated directly via TDNRG and those calculated indirectly via the self-energy function, we observe the same initial and final state satellite peaks, at the expected positions, but with a slight difference in their magnitudes (being slightly sharper within the latter approach as discussed already in Sec.~\ref{subsec:self-energy-errors}). The oscillations around $|T|\leq 1/\Delta$ and their decay, in the interacting case, are similar to those in the noninteracting case
except that the decay is stronger due to the additional broadening coming from the self-energy. In order to see this explicitly, consider the case of small  $T$ close enough to $0^+$ such that we can assume that the self-energy is time independent in the range between $0$ and $T$.
Then from Eq.~\ref{eq:solution1} we have for $T>0$,
\begin{widetext}
\begin{align}
G^{\sigma}_d(T>0,\omega)&\approx e^{2i[\omega-\varepsilon_d^f-\Gamma^f(\omega)-\Sigma^{\sigma}(\omega)]T}g(\omega)-2i\int_0^T e^{2i[\omega-\varepsilon_d^f-\Gamma^f(\omega)-\Sigma^{\sigma}(\omega)](T-T_1)}dT_1\nonumber\\
&=e^{2i[\omega-\varepsilon_d^f-\Gamma^f(\omega)-\Sigma^{\sigma}(\omega)]T}g(\omega)+\frac{1- e^{2i[\omega-\varepsilon_d^f-\Gamma^f(\omega)-\Sigma^{\sigma}(\omega)]T}}{\omega-\varepsilon_d^f-\Gamma^f(\omega)-\Sigma^{\sigma}(\omega)}.\label{eq:expandsolution1}
\end{align}
\end{widetext}
The form of this equation is similar to Eq.~(\ref{eq:nonint1}) for the noninteracting case at $T>0$, namely the Green's function is made up of oscillatory and decay terms
of the form $e^{2i[\omega-\varepsilon_d^f-\Gamma^f(\omega)-\Sigma^{\sigma}(\omega)]T}=e^{2i[\omega-\varepsilon_d^f-\mathrm{Re}[\Sigma^{\sigma}(\omega)]]T}e^{-2T(\Delta -\mathrm{Im}[\Sigma_{\sigma}(\omega)])}$. The appearance of the self-energy in the above expressions explains the reduction of the oscillatory
contributions relative to those in the noninteracting case. Similar arguments can be used to explain the reduced magnitude of the oscillatory contributions at negative times $-1/\Delta \lesssim T \lesssim 0$.

The time-evolution at the low-energy scales in the interacting case is obviously different from that in the noninteracting case since there is no low-energy scale in the latter. In both the spectral functions calculated either directly from the TDNRG or indirectly via the self-energy function, we observe a clear decay of the initial-state Kondo peak around $\omega=0$ at $T=-1/T^{\rm i}_{\rm K}\approx -800/T_{\rm K}$, and the formation of the final-state Kondo peak at $T=1/T_{\rm K}$. At intermediate times $-800/T_{\rm K}<T<1/T_{\rm K}$, the distributions of spectral weight around $\omega=0$ are different in the two cases, a result of the different ways used to calculate the spectral functions. As discussed above in Sec.~\ref{subsec:self-energy-errors}, the calculation of the self-energy function as a ratio of $F(\omega,T)$ and $G(\omega,T)$ gives a more accurate value than that obtained by inverting $G(\omega,T)$. Therefore, the calculation with the self-energy function gives improved results over the direct approach via TDNRG. Nonetheless, both the calculations with and those without the closed Green's function expression show the same overall picture for the time-evolution of the spectral function.

\begin{figure}%
\includegraphics[width=0.48\textwidth]{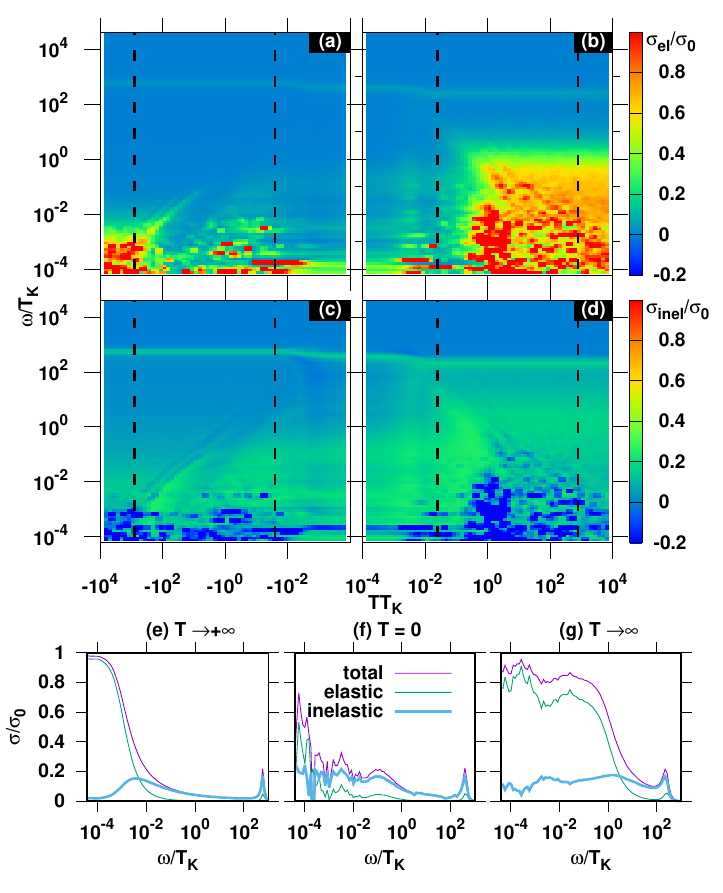}
\caption{(a) and (b) The elastic scattering rate is calculated as $\sigma_{\rm el}=\frac{2\pi}{v_k}\sum_{k'}\delta(\varepsilon_k-\varepsilon_{k'})|\mathbf{T}_{kk'\sigma}(\omega,T)|^2$. (c) and (d) The inelastic scattering rate is given by $\sigma_{\rm inel}=\sigma_{\rm total}-\sigma_{\rm el}$ with $\sigma_{\rm total}=-\frac{2}{v_k}\mathrm{Im}[\mathbf{T}_{kk\sigma}(\omega)]$ with $v_k$ being  the velocity of conduction electrons with wavenumber $k$. The T-matrix is assumed to be $k$-independent, as in the equilibrium case, and  $\mathbf{T}(\omega,T)=|V|^2G(\omega,T)$,  with $G(\omega,T)$ as shown Figs.~\ref{fig:compare1}(e) and (f). The line-cuts of the total, elastic and inelastic scattering rates at $T=-\infty$, $0$, and $+\infty$ are shown in  (e), (f), and (g). All the represented scattering rates are normalized by using the formula $\sigma/\sigma_0$ with $\sigma_0=\frac{2}{\pi\rho v_k}$.
Quench and TDNRG parameters are as in Fig.~\ref{fig:compare3}.
}%
\label{fig:compare4}%
\end{figure}

In Sec~\ref{subsec:self-energy-errors} we briefly mentioned that the Friedel sum rule, $\pi\Delta A(\omega=0,T)=1$ \footnote{In referring to ``violations'' of this equation below, the quotation marks remind the reader that this equation is not an exact result for general times.}, is well satisfied at $T=\pm\infty$, more so at $T=-\infty$ than at $T=\infty$.  Since the sum rule is an equilibrium Fermi-liquid theory result \cite{Hewson1997}, any violation in the above limits indicates the degree of departure from equilibrium in these limits. In this light, the observed small violation at $T=\infty$ is consistent with the imperfect thermalization due to the use of a Wilson chain\cite{Rosch2012} and the resultant small heating at long times \cite{Nghiem2020}. In contrast, for $T=-\infty$, these issues do not arise and the Friedel sum rule was observed to be satisfied to high accuracy
[Figs.~\ref{fig:compare3}(e) and \ref{fig:compare3}(f)]. For general times $T$, we do not expect that $\pi\Delta A(\omega=0,T)=1$ will hold as inelastic scattering processes in the nonequilibrium state will result in a finite imaginary part to the self-energy, resulting in a violation of the above. Here,
we shed more light on this by considering the elastic and inelastic scattering rates \cite{Zarand2004,Bulla2008} with the (approximate) assumption that the T-matrix $\mathbf{T}(\omega,T)=|V|^2G(\omega,T)$. In Figs.~\ref{fig:compare4}(a) and \ref{fig:compare4}(b), we show the elastic scattering rates at positive and negative times, while in
Figs.~\ref{fig:compare4}(c) and \ref{fig:compare4}(d) we show the inelastic ones. One sees that the elastic scattering rate at $\omega$ corresponding to the satellite peak $|\varepsilon_d^i|=U^i/2$ and $|\varepsilon_d^f|=U^f/2$ is quite small in comparison to the inelastic scattering rate. These are incoherent peaks, so a large inelastic scattering is expected. At low frequencies, close to the Fermi level, the elastic scattering rate is close to $1$ at large negative and positive times, while at short times, it is significantly reduced. In contrast, at low energies, the inelastic scattering rate is reduced, vanishing in the
initial state $T\to-\infty$ and reduced significantly also in the final state $T\to +\infty$. These trends are shown in more detail in the line cuts in Figs.~\ref{fig:compare4}(e)-\ref{fig:compare4}(g) for the cases $T=-\infty$, $0$, and $+\infty$. We see that at $T=-\infty$ the scattering rates recover the expected ones for the equilibrium initial-state with the elastic scattering rate as large as the total scattering rate as $\omega \to 0$ while the inelastic scattering rate vanishes (hence the Friedel sum rule is well satisfied in this limit).  At $T=+\infty$, a significant inelastic scattering rate remains for $\omega \to 0$, which we attribute to heating in the imperfectly thermalized final state \cite{Rosch2012,Nghiem2020}. This also suggests a reason for the small violation of the Friedel sum rule in the limit $T\to\infty$. The ``violation'' of $\pi\Delta A(\omega=0,T)=1$, however, was seen to be largest at short times $T=0$ [see Fig.~\ref{fig:compare3}(f)]. This is consistent also with the significant inelastic scattering rate seen in Fig.~\ref{fig:compare4}(f) for $T=0$. Physically,  one expects that during the large spectral weight rearrangement at short times $|T|\lesssim 1/\Delta$,  this will be accompanied by a significant amount of inelastic scattering, even at low energies $|\omega|\lesssim 1/T_{\rm K}$, thus resulting  in the observed ``violation'' of the Friedel sum rule at these times.

\hn{In the noninteracting case, one can see directly from the analytic equations (\ref{eq:nonint1}) and (\ref{eq:nonint2}) for the Green's function that the Friedel sum rule cannot hold at finite times. For example, at $T=0$, the spectral function at $\omega=0$ reads
\begin{align}
  \pi\Delta A(\omega=0,T=0) &=\frac{\Delta^2}{(\varepsilon_d^i+\varepsilon_d^f)^2/{4}+\Delta^2}\label{eq:fsr-nonint}.
\end{align}
The occupation $n_d(T<0)=n_d^i$ is constant before the quench and we have $\sin^2(\pi n_d(T=0)/2)=\Delta^2/({\varepsilon_d^i}^2+\Delta^2)$. Thus the Friedel sum rule $\pi\Delta A(\omega=0,T=0)=\sin^2(\pi n_d(T=0)/2)$ does not hold at $T=0$ (except for $\varepsilon_d^i=\varepsilon_d^f$, i.e., in the absence of a quench). For such a noninteracting system, the violation of the Friedel sum rule is caused by decoherence effects in the nonequilibrium transient state.}

\section{Conclusions}
\label{sec:conclusion}
In summary, we generalized the self-energy method for calculating equilibrium spectral functions for the Anderson impurity model via the correlation self-energy \cite{Bulla1998} to the time-dependent case of a quench from an initial to a final state. For this purpose, we obtained closed expressions for the retarded two-time Green's function for both positive and negative Wigner times. The closed expressions give additional analytic insights into the time-evolution of the spectral function of the Anderson impurity model, complementing the detailed numerical evaluations. We further demonstrated that the time-dependent spectral function calculated via the
correlation self-energy  as a ratio of $F(\omega,T)$ and $G(\omega,T)$ shows improved accuracy over that calculated
directly within the TDNRG. Additional insight into the small violation of the Friedel sum rule at $T\to +\infty$, and
into the more significant ``violation'' at short times $|T|\lesssim 1/\Delta$ was gained by estimating the inelastic scattering rate. While the former is intrinsic and due to the use of a Wilson chain in the NRG calculations, the ``violation'' at short times is a real effect due to the increased inelastic scattering expected in the process of the large spectral weight rearrangement occurring on these time scales (i.e., no Friedel sum rule is expected to hold at these times). The technique developed here holds for both zero and finite temperatures and is applicable within any impurity solver that can access the time-evolution of the single-particle Green's function $G$ and the \hn{higher-order correlation} functions $F$ and $\tilde{F}$ [Eqs.~(\ref{eq:F}) and (\ref{eq:F-tilde})]. For example, the latter are easily calculable within the CT-QMC approach and have already been
used as estimators for the correlation self-energy in the equilibrium case \cite{Hafermann2012}. A further extension to the time-dependent case therefore appears feasible within this method. Other impurity solvers for  time-evolution, where the present approach may prove useful, include the time-dependent non-crossing approximation \cite{Shao1994a}, the TD-DMRG approach \cite{Daley2004,White2004}, \hn{and the real-time bold-line Monte Carlo method \cite{Cohen2014a,Cohen2014b}}. Finally, we note that the self-energy method, considered in this paper for the Anderson impurity model, is not restricted to just this model, but can be applied to a wide range of  other quantum impurity systems. For example, with a suitable generalization of the self-energy function, it can be applied to the Anderson-Holstein model \cite{Jeon2003}.

\begin{acknowledgments}
H. T. M. N. wishes to thank Doan Duy Hai and Pham Tien-Lam for a number of stimulating discussions. H. T. M. N. acknowledges the support by Vietnam National Foundation for Science and Technology Development (NAFOSTED) under Grant No. 103.01-2018.12.
We acknowledge support by the Deutsche Forschungsgemeinschaft via  Research Training Group 1995  and supercomputer support by the John von Neumann institute for Computing (J\"ulich).
\end{acknowledgments}

\appendix

\section{Wide-band limit}
\label{sec:wbl}
Applying the Fourier transform to Eqs.~(\ref{eq:Gre1b}) and (\ref{eq:Gre3b}), we have
\begin{align}
&\Big[\omega+i\frac{\partial}{2\partial T}-\varepsilon^f_d\Big]G^{\sigma}_d(T,\omega)=1+\sum_kV^f_{dk}G^\sigma_{kd}(T,\omega)+U^fF^\sigma_{d}(T,\omega),\label{eq:Gre1bw}\\
&\Big[\omega+i\frac{\partial}{2\partial T}-\varepsilon_{k\sigma}\Big]G^{\sigma}_{kd}(T,\omega)=V_{kd}^fG^\sigma_{d}(T,\omega),\label{eq:Gre3bw}.
\end{align}
Solving the latter, yields
\begin{align}
&G^{\sigma}_{kd}(T,\omega)=\Big[1+\frac{i}{\omega-\varepsilon_{k\sigma}}\frac{\partial}{2\partial T}\Big]^{-1} \frac{V_{kd}^f}{\omega-\varepsilon_{k\sigma}}G^\sigma_{d}(T,\omega).\nonumber
\end{align}
Hence,
\begin{align}
&\sum_kV^f_{dk}G^{\sigma}_{kd}(T,\omega)\nonumber\\
&=\sum_k\Big[1+\frac{i}{\omega-\varepsilon_{k\sigma}}\frac{\partial}{2\partial T}\Big]^{-1} \frac{|V_{kd}^f|^2}{\omega-\varepsilon_{k\sigma}}G^\sigma_{d}(T,\omega)\\
&=\sum_k\Big[1+\sum_{n=1}^{\infty}\frac{(-i)^n}{(\omega-\varepsilon_{k\sigma})^n}\frac{\partial^n}{2^n\partial T^n}\Big] \frac{|V_{kd}^f|^2}{\omega-\varepsilon_{k\sigma}}G^\sigma_{d}(T,\omega).\nonumber
\end{align}
Replacing $\sum_k \frac{|V_{kd}^f|^2}{(\omega-\varepsilon_{k\sigma})^{n}}$ by $\int_{-D}^D d\varepsilon \frac{|V^f|^2\rho}{(\omega-\varepsilon)^{n}}$ \cite{Anders2008a}, and using the wide-band limit, we have that the integral reduces to just the sum of the residues resulting from the contour integration. Since $\mathrm{Res}_{z=0} (1/z)=1$ and $\mathrm{Res}_{z=0} (1/z^n)=0$ for $n>1$, we have
\begin{align}
&\sum_kV^f_{dk}G^{\sigma}_{kd}(T,\omega)=-i\pi |V^f|^2\rho G^\sigma_{d}(T,\omega) = -i\Delta^f G^\sigma_{d}(T,\omega).
\end{align}
Substituting this into Eq.~(\ref{eq:Gre1bw}), we have Eq.~(\ref{eq:ODE1}). A similar derivation holds
for Eq.~(\ref{eq:ODE2}) starting with Eqs.~(\ref{eq:Gre2b}) and (\ref{eq:Gre4b}).

\section{ODE solvers}
\label{sec:ode}

Even though the Green's function has been expressed in a closed analytic form in Eqs.~(\ref{eq:solution1}) and (\ref{eq:solution2}), its evaluation 
can only be carried out numerically since the self-energy function has no closed 
analytic expression. In the calculation of the time-dependent spectral function, solving the ODEs~(\ref{eq:ODE1}) and (\ref{eq:ODE2}) numerically is just equivalent to calculating the 
integrals in Eqs.~(\ref{eq:solution1}) and (\ref{eq:solution2}). Both the ODEs take the form 
\begin{align}
G'(T)=f[T,G(T)].
\end{align}
Integrating the equation from $T_n$ to $T_{n+1}$, we have
\begin{align}
G(T_{n+1})-G(T_{n})=\int_{T_{n}}^{T_{n+1}}f[T,G(T)]dT.
\end{align}

Using the left-hand rule, we have the formula for the explicit Euler method,
\begin{align}
G(T_{n+1})-G(T_{n})=hf[T_n,G(T_n)],
\end{align}
with $h=T_{n+1}-T_{n}$, where the new value $G(T_{n+1})$ is given explicitly in 
terms of the old value $G(T_{n})$. 

Using the right-hand rule, we have the formula for the implicit Euler method,
\begin{align}
G(T_{n+1})-G(T_{n})=hf[T_{n+1},G(T_{n+1})],
\end{align}
where the right-hand side is evaluated at the new $G(T_{n+1})$ value, therefore we have to solve this equation for the unknown $G(T_{n+1})$.

Using the trapezoidal rule, we have the formula for the trapezoidal method,
\begin{align}
G(T_{n+1})-G(T_{n})=\frac{h}{2}\{f[T_{n+1},G(T_{n+1})]+f[T_n,G(T_n)]\}.
\end{align}
Since $G(T_{n+1})$ appears on both sides of this equation, we also have to solve the equation for $G(T_{n+1})$. Therefore,
the above can be also considered as an implicit method. 

One can judge the methods for solving ODEs from two criteria: the accuracy and the stability. Both the explicit and implicit Euler methods have the first order of local accuracy, while the trapezoidal method has the second order of local accuracy \cite{Ascher2011}. For the stability, the explicit Euler method gives unstable results, while the trapezoidal method is stable, although less stable than the implicit Euler method. A detailed discussion of the stability of the different methods will be given in the following sections.

\subsection{Explicit Euler method}
Since $G(T)$ is a complex function,  we have to solve the system of two ODEs. We have Eq.~(\ref{eq:ODE2}) equivalent to
\begin{align}
u_{n+1}=u_n+2hM_nv_n+2hN_nu_n\\
v_{n+1}=v_n+2h-2hM_nu_n+2hN_nv_n
\end{align}
with $u_{n}=u(T_n)=\mathrm{Re}[G(T_n)]$, $v_{n}=v(T_n)=\mathrm{Im}[G(T_n)]$, $M_n=\omega-\varepsilon_d^i-\mathrm{Re}[\tilde{\Sigma}(T_n)]$, and $N_n=\Delta-\mathrm{Im}[\tilde{\Sigma}(T_n)]$. From these equation, we can calculate $u_n$ and $v_n$ recursively with the known $u_0$ and $v_0$. However, the explicit Euler method does not work in the long-time range since $|u_n|$ and $|v_n|$ diverge as $|h|\to \infty$. 

\subsection{Implicit Euler method}
Using the formula for the implicit Euler method, we have Eq.~(\ref{eq:ODE2}) equivalent to
\begin{align}
u_{n+1}=u_n+2hM_{n+1}v_{n+1}+2hN_{n+1}u_{n+1}\\
v_{n+1}=v_n+2h-2hM_{n+1}u_{n+1}+2hN_{n+1}v_{n+1}.
\end{align}
Then
\begin{align}
u_{n+1}=\frac{(1-2N_{n+1}h)u_n+2hM_{n+1}(v_{n}+2h)}{(1-2hN_{n+1})^2+4h^2M_{n+1}^2}\\
v_{n+1}=\frac{(1-2N_{n+1}h)(v_{n}+2h)-2hM_{n+1}u_{n}}{(1-2hN_{n+1})^2+4h^2M_{n+1}^2}.
\end{align}
As $|h|\to \infty$,
\begin{align}
u_{n+1}\to\frac{M_{n+1}}{N_{n+1}^2+M_{n+1}^2}\\
v_{n+1}\to\frac{-N_{n+1}}{N_{n+1}^2+M_{n+1}^2}.
\end{align}
which does not depend on the earlier value $u_n$ and $v_n$. Since $|h|\to \infty$ , $M_{n+1}=\omega-\varepsilon^i_d-\mathrm{Re}[\tilde{\Sigma}(T\to -\infty)]$ and $N_{n+1}=\Delta-\mathrm{Im}[\tilde{\Sigma}(T\to -\infty)]$. We can easily prove that $\tilde{\Sigma}(T\to -\infty)$ is the self-energy of the initial state; therefore the method gives the result converged in one single step. This proves the stability of the method.

\subsection{Trapezoidal method}
Using the formula for the trapezoidal method, we have Eq.~(\ref{eq:ODE2}) equivalent to

\begin{align}
u_{n+1}=u_n&+h(M_nv_n+M_{n+1}v_{n+1})\nonumber\\
&+h(N_nu_n+N_{n+1}u_{n+1})\\
v_{n+1}=v_n&+h-h(M_nu_n+M_{n+1}u_{n+1})\nonumber\\
&+h(N_nv_n+N_{n+1}v_{n+1}).
\end{align}
Then
\begin{widetext}
\begin{align}
u_{n+1}=\frac{(1-N_{n+1}h)(u_n+hM_nv_n+hN_nu_n)+hM_{n+1}(v_{n}+2h-hM_nu_n+hN_nv_n)}{(1-hN_{n+1})^2+h^2M_{n+1}^2}\\
v_{n+1}=\frac{(1-N_{n+1}h)(v_{n}+2h-hM_nu_n+hN_nv_n)-hM_{n+1}(u_{n}+hM_nv_n+hN_nu_n)}{(1-hN_{n+1})^2+h^2M_{n+1}^2}.
\end{align}
As $|h|\to \infty$,
\begin{align}
u_{n+1}\to\frac{2M_{n+1}-u_n(N_{n+1}N_n+M_{n+1}M_n)-v_n(N_{n+1}M_n-M_{n+1}N_n)}{N_{n+1}^2+M_{n+1}^2}\\
v_{n+1}\to\frac{-2N_{n+1}-v_n(N_{n+1}N_n+M_{n+1}M_n)+u_n(N_{n+1}M_n-M_{n+1}N_n)}{N_{n+1}^2+M_{n+1}^2}.
\end{align}
\end{widetext}
which depend on the earlier values $u_n$ and $v_n$. Then, in contrast to the implicit Euler method, the trapezoidal method does not give the result converged in one single step. Only when $T_n$ is large enough such that $M_{n+1}=M_n$ and $N_{n+1}=N_n$, do we have
\begin{align}
u_{n+1}\to\frac{2M_{n+1}}{N_{n+1}^2+M_{n+1}^2}-u_n\\
v_{n+1}\to\frac{-2N_{n+1}}{N_{n+1}^2+M_{n+1}^2}-v_n.
\end{align}
That means the trapezoidal method is less stable than the implicit Euler method, even though, although still converged at large $|T|$.

\subsection{Comparison}
\begin{figure}%
\includegraphics[width=0.48\textwidth]{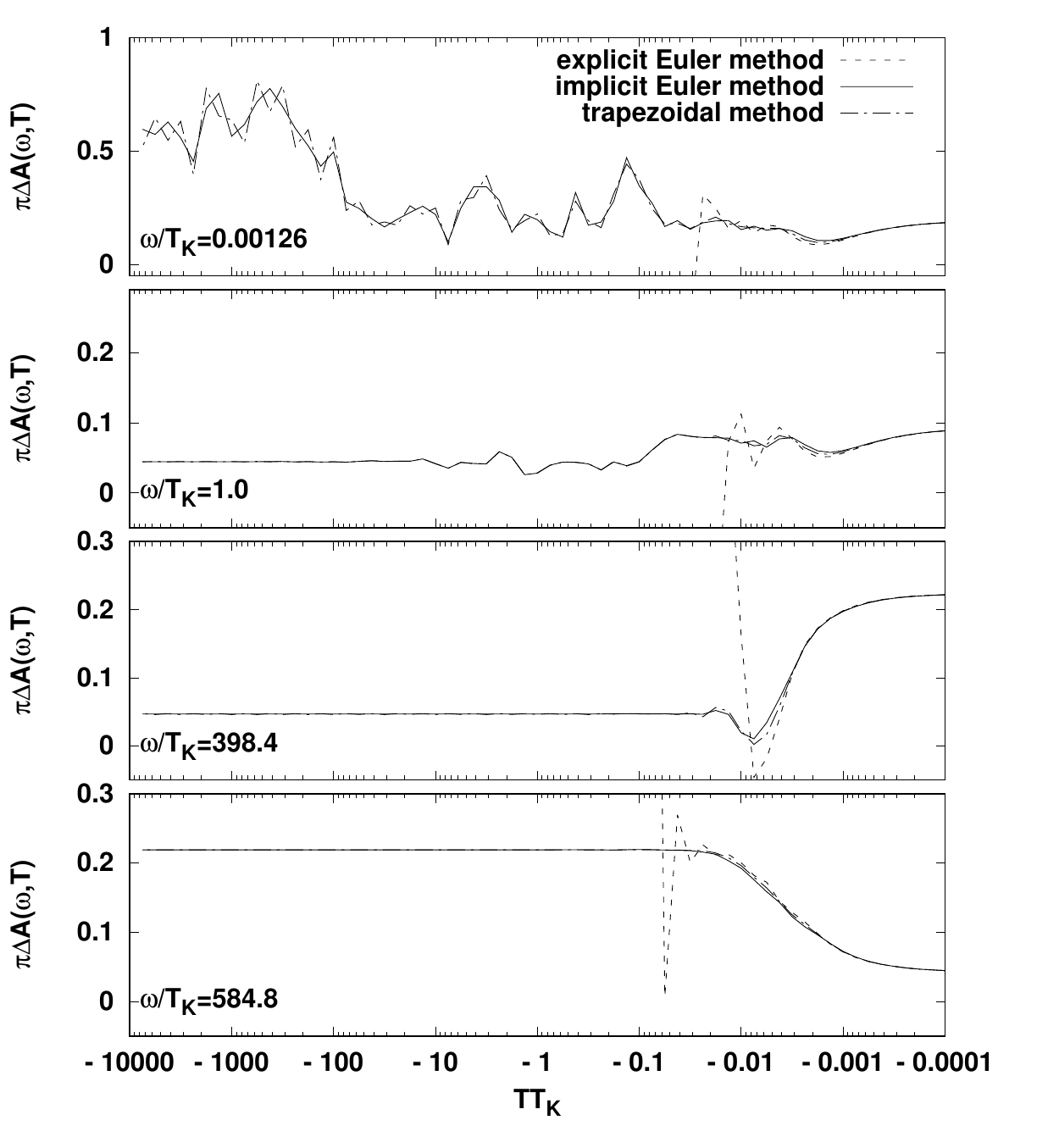}
\caption{Time-evolution of the normalized spectral function at several frequencies $\omega$ calculated with the implicit Euler method (solid line), the explicit Euler method (dashed line), and the trapezoidal method (dot-dashed line).}
\label{fig:view}%
\end{figure}

The results of the spectral function $A(\omega,T)=-\mathrm{Im}[G(\omega,T)]/\pi$ in Fig.~\ref{fig:view} show the detailed properties of each method in solving the ODE.~(\ref{eq:ODE2}). For times $T$ close to zero, the step size is small due to the use of a logarithmic time-grid about $T=0$, hence the three solvers give similar results. For times $T$ further away from zero, the step size (also due to the use of a logarithmic grid for the time axis) is larger, the explicit Euler method starts to become 
unstable and eventually the results diverge for large enough $T$. The implicit Euler method and the trapezoidal method give similar results for the time-evolution, with only small differences in intensities at large negative-times and in the small frequency range. Therefore both of these two methods are good in giving stable results. The results shown in Figs.~\ref{fig:compare1}-(e) and (f) are calculated with the implicit Euler method.

\hn{
\section{Analytic vs TDNRG for the noninteracting case}
\label{sec:noninteracting}

In order to illustrate more clearly which features in the spectral functions are a result of the TDNRG approach and which features are intrinsic to the time-dependence, we  compare in Fig.~\ref{fig:noninteracting} the spectral functions calculated using the analytic expressions in Eqs.~(\ref{eq:nonint1} and \ref{eq:nonint2}) at several specified times [solid curves are line cuts from Figs.~\ref{fig:compare1}(a) and \ref{fig:compare1}(b)] with those calculated directly from the TDNRG (solid circles). The comparison shows that the TDNRG method overbroadens and reduces the magnitude of the high-energy features in the spectral function. This is due to the
lower resolution at high energies of the logarithmic grid, which is used in both TDNRG and NRG, i.e., ultimately this is due to the broadening scheme used in both NRG and TDNRG. On the other hand, the regions of negative-spectral density, which can be seen for time ranges corresponding to large spectral weight rearrangements, i.e., when
$\varepsilon_d^i\to \frac{\varepsilon_d^i+\varepsilon_d^f}{2}$ and when $\frac{\varepsilon_d^i+\varepsilon_d^f}{2}\to \varepsilon_d^f$, are a common feature of both the TDNRG and
the analytic results. For the interacting case, the latter conclusion was also demonstrated for the time evolution of the Kondo resonance following a quench \cite{Nghiem2017}. 

\begin{figure}%
\includegraphics[width=0.48\textwidth]{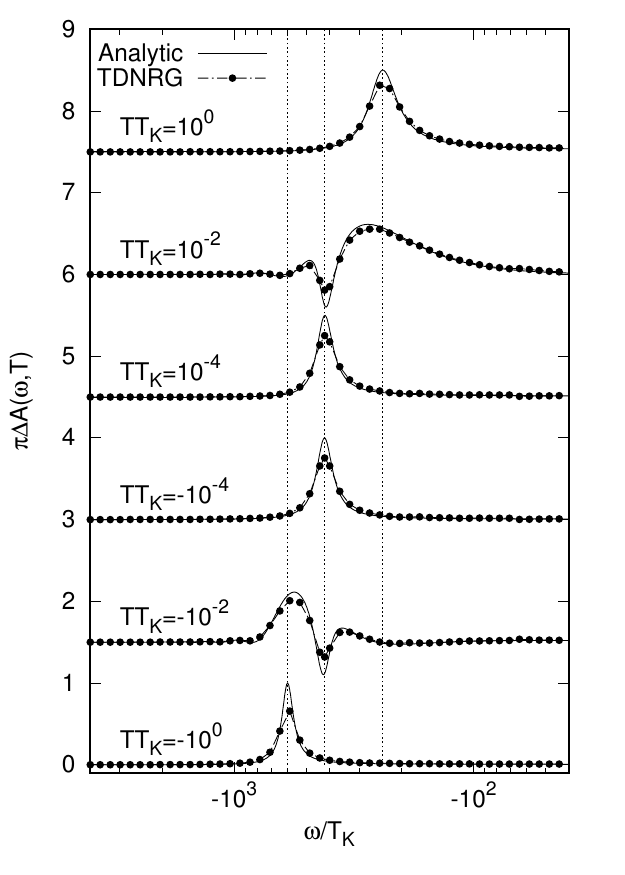}
\caption{Spectral function at several times $TT_{\rm K}$, as indicated, and shifted vertically by increments of $1.5$ for clarity, for the noninteracting case $U^i=U^f=0$ calculated analytically from Eqs.~(\ref{eq:nonint1}-\ref{eq:nonint2}) (solid curves) and numerically within TDNRG (filled circles). Vertical dashed lines, from left to right, show the initial ($\varepsilon_{d}^i$), intermediate ($\frac{\varepsilon_d^i+\varepsilon_d^f}{2}$) and final state ($\varepsilon_d^f$) local level positions. Local-level quench and TDNRG parameter as in Figs.~\ref{fig:compare1}(a)-\ref{fig:compare1}(b).}
\label{fig:noninteracting}%
\end{figure}
}
\bibliography{noneq-nrg}
\end{document}